\PassOptionsToPackage{bookmarks=false}{hyperref}
\documentclass[sigconf]{acmart}

\usepackage{booktabs} 
\usepackage{enumitem}

\setcopyright{acmcopyright}

\begin{document}
	
\title{Code Review Comments: Language Matters 
}

	\author{Vasiliki Efstathiou and Diomidis Spinellis}	
	\email{{vefstathiou, dds}@aueb.gr}  
    \affiliation{%
	\institution{Department of Management Science and Technology \\ Athens University of Economics and Business}
	\city{Athens} 
	\state{Greece} 
}

\copyrightyear{2018} 
\acmYear{2018} 
\acmISBN{978-1-4503-5662-6/18/05} 
\acmConference[ICSE-NIER'18]{40th International Conference on Software Engineering: New Ideas and Emerging Results Track}{May 27-June 3, 2018}{Gothenburg, Sweden}
\acmBooktitle{ICSE-NIER'18: 40th International Conference on Software Engineering: New Ideas and Emerging Results Track, May 27-June 3, 2018, Gothenburg, Sweden}
\acmPrice{15.00}
\setcopyright{acmcopyright} 
\acmDOI{10.1145/3183399.3183411}

\begin{abstract}
Recent research provides evidence that effective communication in collaborative software development has significant impact on  the software development lifecycle.  
Although related qualitative and quantitative studies point out textual characteristics of well-formed  messages, the underlying semantics of the intertwined linguistic structures still remain largely misinterpreted or ignored. Especially, regarding quality of code reviews the importance of thorough feedback, and explicit rationale is often mentioned but rarely linked with related linguistic features. As a first step towards addressing this shortcoming, we propose grounding these studies on theories of linguistics. 
We particularly focus on linguistic structures of coherent speech and explain how they can be exploited in practice. We reflect on related approaches and examine through a preliminary study on four open source projects, possible links between existing findings and the directions we suggest for detecting textual features of useful code reviews.

\end{abstract}



\maketitle

\section{Introduction}
Archived communications from developer interactions provide a rich source of information which, 
under appropriate analysis and interpretation, may reveal interesting facts related to   
practices, decisions and team dynamics in a collaborative development environment. 
Software peer review is a QA practice widely adopted by open source and commercial projects 
 \cite{rigby2013convergent} where, besides finding defects, it offers additional benefits such as knowledge transfer across members of a development team \cite{bacchelli2013expectations}. 
To this end, a number of qualitative studies have been conducted towards identifying factors that influence code review usefulness including, among others, the textual characteristics of the review comments \cite{bacchelli2013expectations}, \cite{rigby2013convergent}, \cite{kononenko2016code}. Other studies have combined empirical and text mining methods, to identify linguistic aspects of constructive code review comments  \cite{bosu2015characteristics},  \cite{rahman2017predicting}. However, the existing studies are limited to superficial textual characteristics such as stop word ratio and sentence  length, with no semantic interpretation. 
Current research for deriving semantics from natural language exchanged by developers is largely directed towards the extraction of behavioural aspects which may only indirectly affect the progress of a project 
\cite{guzman2013towards}, \cite{ortu2015bullies}. 
Related studies applying sentiment analysis research with software engineering data are heavily based on 
off-the-shelf tools, which are trained in different contexts, such as product or movie reviews and 
may therefore not apply to other domains.  
In addition, although in sentiment analysis research the objective is to mine opinions related to specific artefacts and the valence of emotions triggered by these artefacts, 
adaptations in software engineering arbitrarily assume a {\it unidirectional} causal relationship from negative team emotions to negative effects on the performance of a collaborative project. That is to say, the effect that a toxic project may have on the dynamics of the development team is not taken into consideration. 
Consequently, negative results regarding the suitability of applying 
state of the art sentiment analysis methods on software engineering research have started to emerge \cite{jongeling2017negative}. Ongoing research works towards filling this gap 
by collecting domain-specific software engineering data \cite{ortu2016emotional}, \cite{mantyla2017bootstrapping}, or performing more fine grained classification of emotions in collaborative software development~\cite{gachechiladze2017anger}. 

In this work we investigate aspects of linguistic semantics that go beyond emotion and opinion valence detection. 
Working on the hypothesis that communication in a collaborative environment primarily concerns conveying factual objective information rather than expressing subjective opinions, we investigate facets in language which may indicate cues of effective collaborative  messages. 
We focus our study on code reviews due to their value in software development and the increasing interest of the community for improved practices  \cite{bacchelli2013expectations}. 
We believe however that the outcomes from the directions we propose are transferable to other  types of communications related to software development and maintenance, such as bug reports, issue tracking comments or informal chats.
\section{Related work}
\label{sec:related}
The study by Rahman et al. \cite{rahman2017predicting} is the one 
that shares objectives closest to ours. The authors acknowledge the 
importance of effective code review comments, point out the lack of 
related automated text mining techniques and attempt to interpret 
findings from  empirical studies into textual properties of comments \cite{bosu2015characteristics}, \cite{kononenko2016code}. 
Taking into account the change-triggering effect of a review as a 
heuristic for usefulness, they extracted a number of features from the 
text of the review comments. 
Their study however is limited to structural characteristics of the text with no consideration of interpreting the underlying semantics. 
Based, for instance, on the finding from Kononenko et al. 
\cite{kononenko2016code} that 
{\it `Clear and thorough feedback is the key attribute of a well-done code review'} they study clarity on the basis of reading ease, defined in terms of word and sentence length. Reading ease and question ratio in comments did not provide significant results in their study. However other dimensions did, 
 with most prominent being the presence of 
 actual source code elements quoted in the review text and the `{\it conceptual similarity}', a metric that captures the proportion of 
vocabulary shared between the review comment and the corresponding modified source code. 
Another factor taken into account was the stop word ratio, which proved to be higher in non-useful comments, albeit with a small effect size.  
Based on the derived textual and other features related to developer expertise, 
 the authors built a model for predicting review comment usefulness. 
 
Similarly, Bosu et al. \cite{bosu2015characteristics}, based on the findings of an interview-based empirical study, built a decision tree classifier for distinguishing between useful and not useful code review comments. They used attributes related to reviewer expertise, change set and revision thread metrics and textual features of the review. The textual features they used comprised a vocabulary of 349 keywords that were frequently found in comments that had been classified as useful through the empirical study. 

In this study we investigate exclusively linguistic information 
contained in the text of the comment. 
The benefit of assessing a review by its comment is that it does 
not require information that can only be 
available upon completion of the review, such as thread and 
change set metrics mentioned in the previous paragraph. 
Looking at the comment text alone, offers opportunities for tools 
classifying a comment at real-time and recommending improvements 
before its submission. 
From the features mentioned in this section the ones that are solely 
language-based and gave promising results are the existence of source 
code elements in the comment \cite{rahman2017predicting} and the set 
of keywords mentioned by Bosu et al. \cite{bosu2015characteristics}. 
\section{Functional semantics of language}
Peer code review is by definition a judgmental process; it involves reviewing and assessing whether a change suggested by a peer is appropriate for merging into the codebase. As such, the related communications have to reflect the reasons why a change needs to be made and consequently, reasons  for accepting or rejecting this change. The need for appropriately justified code review comments has been identified in the literature under the notions of rationale articulation  
\cite{bacchelli2013expectations}, \cite{sutherland2009rationale} or  
thoroughness of the review feedback \cite{kononenko2016code}. 
In this section we propose directions for grounding these findings in longstanding linguistic theories of rhetorics and discourse coherence \cite{mann1988rhetorical}. 
The idea is that there exist implicit relations 
between the sentences of a text so that the content of one sentence might provide, for example,  elaboration  or explanation for the content of another. These relations  
 bind a text together, thus contributing to its overall coherence, and 
they are often made explicit by the speaker through the use of particular cue phrases. 
In Table \ref{tab:keywords} we give some examples of such phrases along with their functional semantics as described by Knott and Dale \cite{knott1994using}.  
The cue phrases in Table \ref{tab:keywords} are examples extracted from a comprehensive vocabulary compiled in by Knott and Dale \cite{knott1994using}, which has served as the basis for deriving sets of more complex lexico-syntactic patterns for mining instances of justifications, analogies and conditional 
statements in text \cite{lippi2015context}, \cite{mochales2011argumentation}, \cite{walton2008argumentation}. These types of structures 
are context-independent and have been studied across diverse domains such as legal reasoning \cite{ashley2013toward}, medical decision support \cite{hunter2012aggregating}, policy making \cite{bench2015using}, product reviews \cite{wyner2012semi} and lately for discourse analysis in social media \cite{dusmanu2017argument}. 
We argue that a similar approach is a promising direction for assessing the linguistic characteristics of useful code reviews. We motivate our approach by revisiting examples 
from  existing work \cite{rahman2017predicting} and provide a preliminary analysis on a set of code reviews from four open source projects \cite{Yang2016MSR}. 
\begin{table}
	{\small
		\begin{tabular}{ll}
			\toprule	
			{\bf Functionality} & {\bf Cue phrases}\\
			\midrule
			Causality & Because, since, as, thus, as a result\\
			Contrast & But, however, although, on the other hand \\
			Exemplification  & For example, for instance \\
			Clarification & In fact, actually, in other words, in short\\
			Similarity & Also, as well, similarly, too\\
			Hypothesis & If, in this case, assuming that, provided that\\ 
			\bottomrule
	\end{tabular}}
	\caption{Examples of coherence relations}
	\label{tab:keywords}
	\vspace{-27pt}
\end{table}
\section{Useful comments revisited }
\label{sec:revisited}
Based on the assumption that stop words carry {\it `very little or no semantics'}, Rahman et al. 
worked on the hypothesis that a high ratio of stop words in text is a characteristic of non-useful code reviews. 
Stop words are words frequently occurring in natural language and although there is no universal list, state of the art vocabularies of stop words include: {\it a, also, as, but, by, so, the, while} among others. 
 In settings such as document retrieval or topic modelling, they are often removed because their presence in a text does not contribute in distinguishing one concrete topic in text from another. However, even though they are topic-agnostic, in the role of coherence cue words they are necessary to construct higher level semantical meaning, 
 which may be important for articulating accurate comments. This is demonstrated in the single  
 concrete example of useful code review comment presented by Bosu et al. \cite{bosu2015characteristics}: 
	 {\it `I don't think we need 2 ways to call \mbox{\texttt{\small get\_partner\_whitelabel\_config}} as \mbox{\texttt{\small market\_id}} is \mbox{\texttt{\small None}} by default}. 
Even though the authors acknowledge that this comment may be useful due to the warranted action it expresses, the textual feature they locate as indicative of usefulness is the presence of source code artefacts. 
The source code itself though would possibly be of little use if it had not been an object of a justification phrase expressed by the term `{\it as}'. However, it is common practice in text mining studies \cite{bosu2015characteristics} to remove such words during  preprocessing. 

Motivated by this example, we investigated the assumption that {\it source code elements appear often in useful comments because they act as references in explanatory phrases}. 
To this end, we studied the inclusion of coherence relations in the sets of words that  frequently appear in close proximity with source code elements in comments. 
More specifically, we examined the {\it collocation} of source code and linguistic coherence elements in the comment messages of code reviews. The objective was to study whether references to source code are often coupled with 
instances of coherence relations. 

For our study we performed a preliminary analysis on publicly available code review data from four open source projects.\footnote{\url{http://kin-y.github.io/miningReviewRepo/}} We selected this data set because it contains a substantial volume of data ($\sim$ 250,000 patches) from the well-known open source projects Eclipse, LibreOffice, AOSP and OpenStack, mined from Gerrit code reviews and organised in a portable database dump \cite{Yang2016MSR}.
In this analysis we used the textual information of review messages provided in the data set. 
 During preprocessing we did not perform stop word removal or stemming. The only elements we removed were signature strings that were not part of the comment message  itself (e.g. `Author-Id', `Signed-off-by'). In addition, we replaced  
 non-natural language elements, such as references to URIs, path directories and source code  with a dummy string specific to each type of replacement and identical across all comments. 
 We studied the comments of each project separately. 
For the occurrences of the source code replacement string  we generated all the bigram  collocations within a window of three words, i.e. all the pairs that contain this string and a word located in the text at a distance of one or two words away, either before or after the string. 
We counted the occurrences of all distinct words that appear in bigram collocations with the source code string and ranked them in order of descending frequency. 
In the ranking we excluded the trivial collocations with article `a(n)' and  with  instances of other source code elements. 
 The set of coherence relations that we used for this analysis was composed of 100 single-word expressions compiled from the vocabulary given by Knott and Dale \cite{knott1994using}. 
 
\begin{figure*}
\centering
\includegraphics[width=0.245\textwidth]{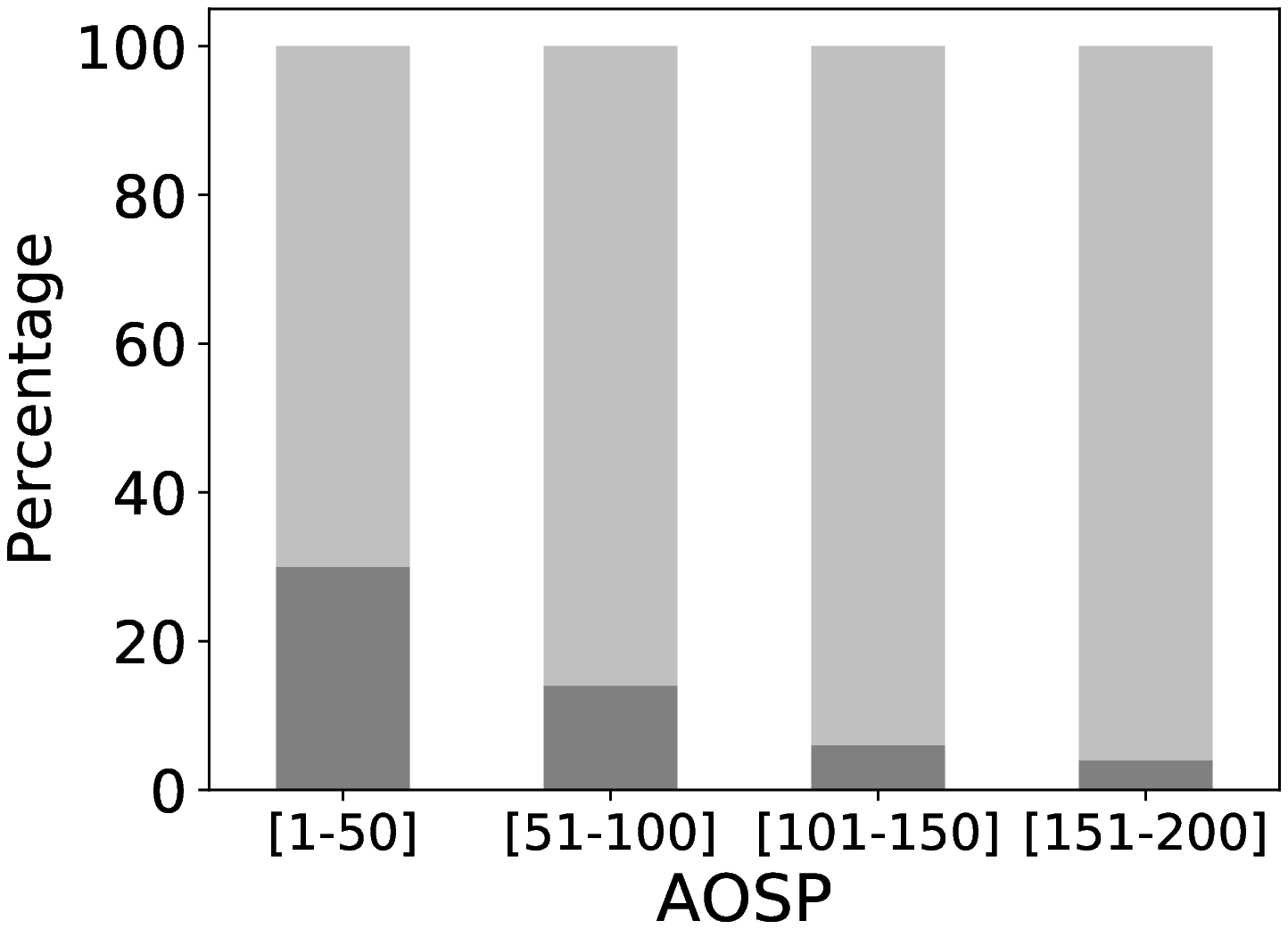}
\includegraphics[width=0.245\textwidth]{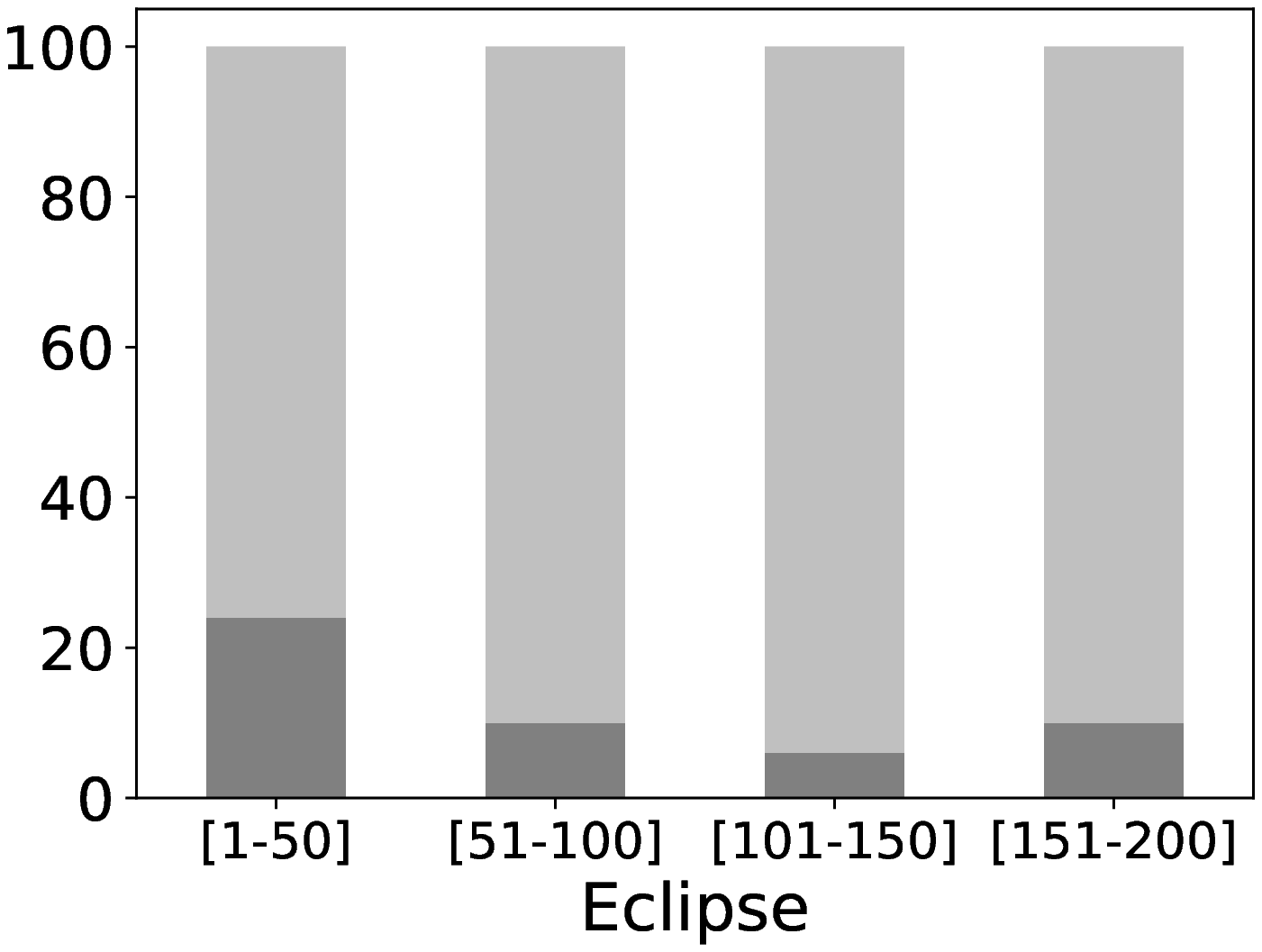}
\includegraphics[width=0.245\textwidth]{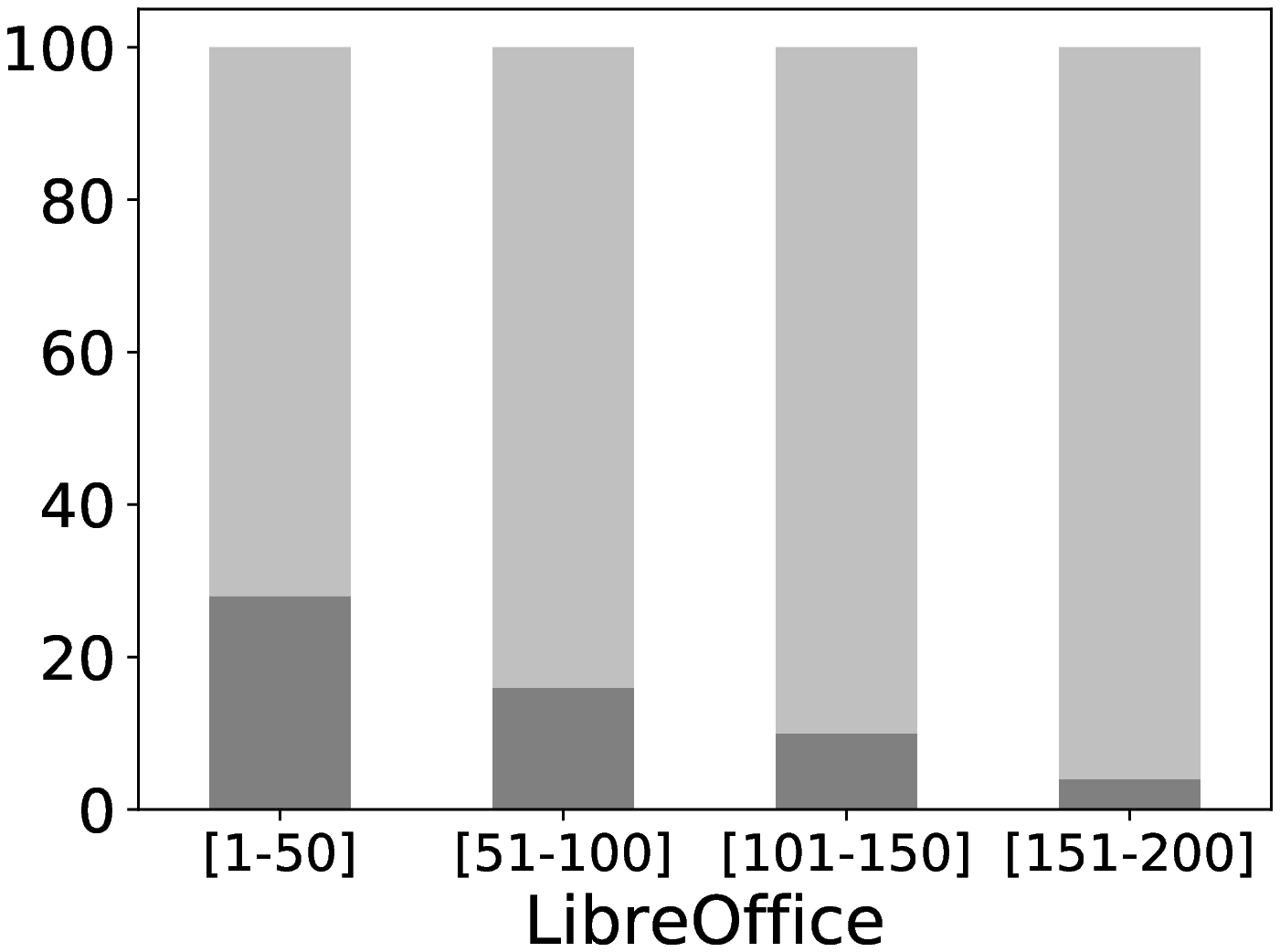}
\includegraphics[width=0.245\textwidth]{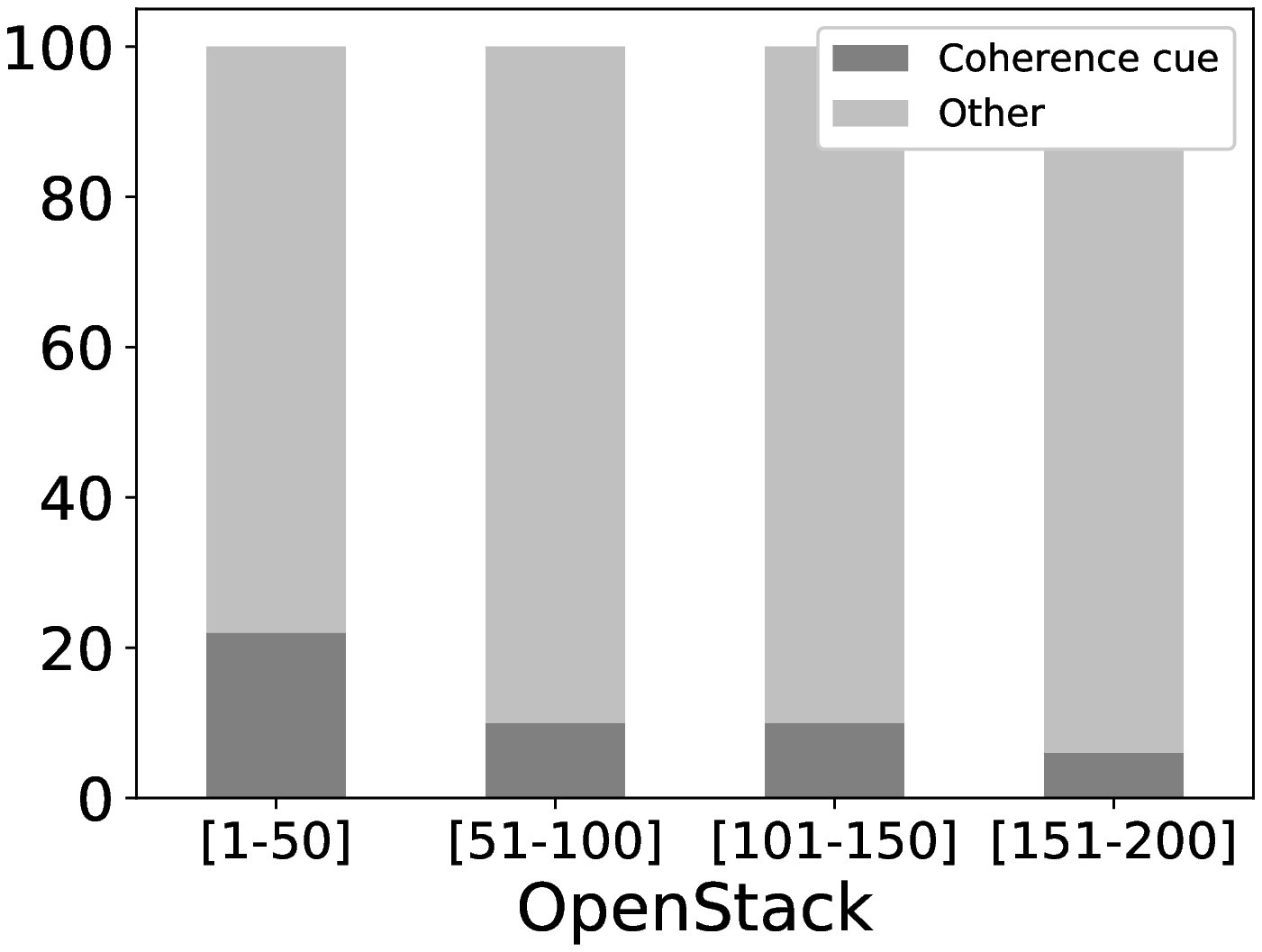}
\vspace{-10pt}
\caption{ Coherence keyword inclusion in  top 200 bigram collocations with source code elements in four different projects.}
\vspace{-14pt}
\label{fig:collocatios}
\end{figure*}

In order to quantify the existence of coherence relations in close proximity with source code in the comments, we computed the inclusion rate of the coherence keyword vocabulary in the top 50, 100, 150 and 200 most frequent source code bigrams in each project. 
 The analysis showed that  
(a) the top 50 most frequent bigram collocations 
contain words that denote some coherence relation at a rate of 22\%\textemdash30\% across the four  projects, and
(b) for bigram collocations of decreasing frequency the inclusion rate of coherence relations tends to decrease. 
These observations, illustrated in Figure \ref{fig:collocatios}, support the intuition that references to source code elements in review comments are often coupled with instances of coherence relations. 
As indicative examples of coherence relations in this study, we report the coherence  keywords  most frequently collocated with source code in the four projects: 
{\it because, after, or, but, so, instead, since, when, only, if, as, and, for, before, now, not, then, also}. 
The coherence keywords that were most frequently found in collocation with source code in {\it all} four projects (i.e. their intersection) are: 
{\it as, if, not, for, so, and, also, instead, when}. 
Note that the keyword {\it `as'} mentioned in the motivating example appears in the 50 most frequent bigrams of all four projects. Specifically, it appears in the following ranks across all source code bigram collocations of each project: 
$23^{rd}$ of 1000 in AOSP,  $17^{th}$ of 400 in Eclipse, $12^{th}$ of 1000 in LibreOffice and $16^{th}$ of 3000 OpenStack.\footnote{We applied a filter of 10 in the reported totals; words that appear less than 10 times in bigrams with source code elements are not considered.}

The results of this study motivate further, fine-grained analyses that will (i) include complex lexico-syntactic patterns of linguistic coherence beyond single-word expressions (ii) investigate the involvement of references other than source code in coherence relations (iii) characterise the presence of coherence relations of diverse functionalities (e.g. causality, exemplification) in review comments. 
We envision methods for automatic evaluation of useful comments that will incorporate features of linguistic coherence.  
\section{Discussion}
\label{sec:discussion}
The software engineering community has successfully adapted 
 natural language processing methods for identifying textual features of useful code reviews. 
However, there is a need for establishing synergies 
with experts who will provide directions for appropriately interpreting these  findings and will identify requirements for further studies. 

Results from different studies may motivate diverse directions of investigation. 
In the previous section we presented a possible interpretation of the reasons why a specific  code review comment may have been annotated as useful, by examining it from the perspective of linguistic coherence. 
Bosu et al. \cite{bosu2015characteristics}, mention among other findings that {\it `verbs of request (e.g., please, should, may be) are more likely to be useful'}. It is worth noting that 
 {\it`should'} and {\it`may'}  belong to the special grammatical category of modal verbs, which are verbs used to express necessity or likelihood. When interpreted appropriately and examined along with the rest of modals as a category they may reveal  aspects of the messages in which they are being used.  
Moreover, in examples of keywords that appear in useful comments, the authors list among others the verbs {\it expand, match, remove, move, fix}, whereas with keywords frequent in non-useful comments they list verbs {\it store, leave, let, keep, work, fail}.  Apparently, the useful comments contain mostly verbs that denote some kind of transformation whereas the non-useful comments contain mostly verbs that denote no change in state. Linguists have determined that syntactic behaviour is a reflection of the underlying semantics and have established results in semantic categorisation of verbs \cite{levin1993english}. 

Motivated by this discussion, here are our recommendations for the effective processing of natural language text in software engineering artefacts: \\	
	\noindent{\it Semantics.} Interpret the meaning of text mining results at the {\it concept} level. Treat words as concepts rather than as plain lexical units; instead of looking at flat lists of results take into account whole conceptual categories in which they may belong.\\
	\noindent{\it Syntax and grammar.} Do not ignore the lexical semantics  hidden in grammatical and syntactical structures. 
	Text preprocessing often removes valuable linguistic information. 
	During empirical studies this information is displayed to annotators who assess the  usefulness of comments presented to them in full text. We recommend that models also be developed using the original full text. \\
	\noindent {\it Synergies.} In order to satisfy the preceding two recommendations we need to consult research outcomes from  expert linguists. To avoid re-inventing the wheel we should establish synergies with communities which have already put theory into practice and can contribute with methods   \cite{li2016neural}, \cite{lippi2015context},  \cite{mochales2011argumentation} and annotated data \cite{bentivogli2017recognizing}, \cite{burchardt2017fate}, \cite{AIFdb}. 
\section{Opportunities}
Specifying concrete linguistic features for useful code reviews 
offers opportunities for advancing the study and practice of code review in diverse dimensions. 

\noindent {\it Recommendations for good practices.} 
Guidelines for efficient communication coupled with explicit linguistic instructions 
 will not only increase comprehension for the receivers of the messages, but also for the transmitters. Prompting developers to explicitly argue over a suggested change forces them to expose the reasoning to themselves first before they post a comment. 
 This practice could lead to the reduction of instances of non-useful messages such as {\it `when a reviewer incorrectly indicates a problem in the code, perhaps due to lack of expertise'}  \cite{bosu2015characteristics}. 

\noindent {\it Tools}. Good practices can be reinforced through the use of tools which, based exclusively on linguistic features, can provide automated immediate feedback for improvements and prevent non-useful comments from being fed into the review pipeline. 
 

\noindent {\it Archived Data.} Code reviews that capture rationale are valuable not only at the project level; they also comprise a rich resource of archived information for later  needs, such as retrieval of design rationale \cite{bacchelli2013expectations}, \cite{sutherland2009rationale}. 
The research ideas we propose contribute towards this direction. 

\noindent {\it Transferable results.} In this study we focused on code reviews. However, the linguistic structures we examined are context-independent. We believe that the proposed approach has potential on a variety of software engineering processes that involve comments not only as part of direct  communications but also as source code artefacts. 
As cited by Kononenko et al.  \cite{kononenko2016code}, for some developers code quality is about the presence of meaningful comments; `{\it comments should tell why not what}'. 
\section{Conclusions}
Code reviews offer rich linguistic information with unexplored directions for research. Text mining studies have provided hints for useful review comments, however it is only with the guidance of expert linguists that these hints can turn into generalised results. 
Likewise, results from empirical studies on useful comments will be best exploited if interpreted by experts into appropriate linguistic cues.
As a first step towards reconciling theory and practice, we proposed adapting linguistic theories of coherence along with text mining for assessing useful code review comments. Our motivation emerged from results in empirical studies that call for thorough well-justified comments 
as well as indicative examples of useful comments from the literature. 
Our preliminary case study gave encouraging results for following this direction. 
\begin{acks}
	The project associated with this work has received funding from the European Union's Horizon 2020 research and innovation programme under grant agreement No 732223. 
\end{acks}

\bibliographystyle{ACM-Reference-Format}
\bibliography{sample-bibliography} 

\end{document}